\documentclass[
 reprint,
 amsmath,amssymb,
 aps,
 pra
]{revtex4-2}

\usepackage{graphicx}
\usepackage{setspace}
\usepackage{braket}
\usepackage{dcolumn}
\usepackage{bm}
\usepackage{hyperref}
\usepackage{cleveref}

\begin{document}
\onehalfspacing

\title{Towards entanglement-enhanced probing \\ of atomic parity violation}

\author{Maxim~Sirotin}
\email{sirotin@mit.edu}
\affiliation{Massachusetts Institute of Technology, Cambridge, MA 02138, USA}

\date{\today}

\begin{abstract}
Atomic parity violation (APV) provides a low-energy probe of the weak interaction between electrons and nuclei, complementary to collider tests of the Standard Model.  Isotope-chain measurements are especially attractive because they test weak-charge scaling while reducing dependence on absolute atomic-structure theory.  We review the APV mechanism, the state of the art in Cs and Yb, and recent trapped-ion, optical-lattice, and molecular proposals.  Motivated by progress in coherent control of atoms, ions, and molecules, we ask a metrological question: given \(N\) probes distributed over an isotope chain, what quantum strategy optimally measures a deviation from Standard Model weak-charge scaling? The optimum is a particular form of a cross-isotope cat state protocol.  We compare this protocol with the standard quantum limit, squeezed-array, same-isotope cat, and discuss extension to recently suggested decoherence-free subspace protocols. We show that entanglement can strongly accelerate statistical averaging, but the ultimate precision is set by APV-specific systematic floors which require careful studies.
\end{abstract}

{
\let\clearpage\relax
\maketitle
}

\section{Introduction: weak interactions in atoms}

Atomic parity violation (APV) is the appearance of parity-odd observables in atomic transitions due to the weak interaction between bound electrons and the nucleus \cite{Nanos2024, Antypas2019PRA}.  Ordinary atomic structure is dominated by electromagnetism, which conserves parity.  Atomic eigenstates therefore have well-defined parity, and electric-dipole transitions connect states of opposite parity.  The weak interaction is different: it violates parity and mixes opposite-parity electronic states.  As a result, a nominally forbidden transition can acquire a tiny electric-dipole amplitude (\Cref{fig:apv_intro}a).

The dominant nuclear-spin-independent (NSI) electron--nucleus weak Hamiltonian can be written
\begin{equation}
H_{\rm NSI}
=
-\frac{G_F}{2\sqrt{2}}Q_W\gamma_5\rho_N(r),
\label{eq:HNSI}
\end{equation}
where \(G_F\) is the Fermi constant, \(Q_W\) is the nuclear weak charge, \(\gamma_5\) is the Dirac matrix that makes the interaction parity odd, and \(\rho_N(r)\) is the normalized nuclear density.  To leading order in the Standard Model,
\begin{equation}
Q_W\simeq -N+Z(1-4\sin^2\theta_W),
\label{eq:weakcharge}
\end{equation}
where \(Z\) and \(N\) are the proton and neutron numbers, and \(\theta_W\) is the weak mixing angle.  Since \(1-4\sin^2\theta_W\approx0.07\), the weak charge is approximately minus the neutron number.  Thus NSI APV primarily probes the electron--neutron weak interaction (\Cref{fig:apv_intro}a).

The weak interaction is short-ranged on atomic scales, so APV is sensitive to the electron wavefunction inside the nucleus.  The weak-induced electric-dipole amplitude between same-parity states \(\ket{i}\) and \(\ket{f}\) may be written schematically as
\begin{equation}
E_{\rm PNC}
=
\sum_n
\left[
\frac{\bra{f}D\ket{n}\bra{n}H_W\ket{i}}{E_i-E_n}
+
\frac{\bra{f}H_W\ket{n}\bra{n}D\ket{i}}{E_f-E_n}
\right],
\label{eq:sum_pnc}
\end{equation}
where \(D\) is the electric-dipole operator, \(H_W\) is the weak Hamiltonian, and the intermediate states \(\ket{n}\) have opposite parity.  This expression shows why APV depends both on nuclear weak charge and on detailed atomic structure.  For heavy atoms the amplitude is enhanced approximately as
\begin{equation}
E_{\rm PNC}\sim Z^3 R(Z\alpha),
\end{equation}
where \(E_{\rm PNC}\) is the parity-nonconserving (PNC) amplitude, \(\alpha\) is the fine-structure constant, and \(R(Z\alpha)\) is a relativistic enhancement factor.  The \(Z^3\) scaling comes from the increased penetration of $s$ electrons into the nuclear region.  Heavy atoms and ions are therefore natural APV platforms.

There is also a nuclear-spin-dependent (NSD) contribution, commonly parametrized by a dimensionless coefficient \(\kappa\).  NSD parity violation (NSD-PV) is present only for isotopes with nonzero nuclear spin and is sensitive to nuclear anapole moments and weak nucleon--nucleon interactions \cite{Wood1997,DeMille2008,Altuntas2018}.  In this paper we focus on NSI APV, because spin-zero even-isotope chains provide the cleanest weak-charge observable.

\subsection{How APV is measured}

\begin{figure*}[htbp]
  \centering
  \includegraphics[width=\textwidth]{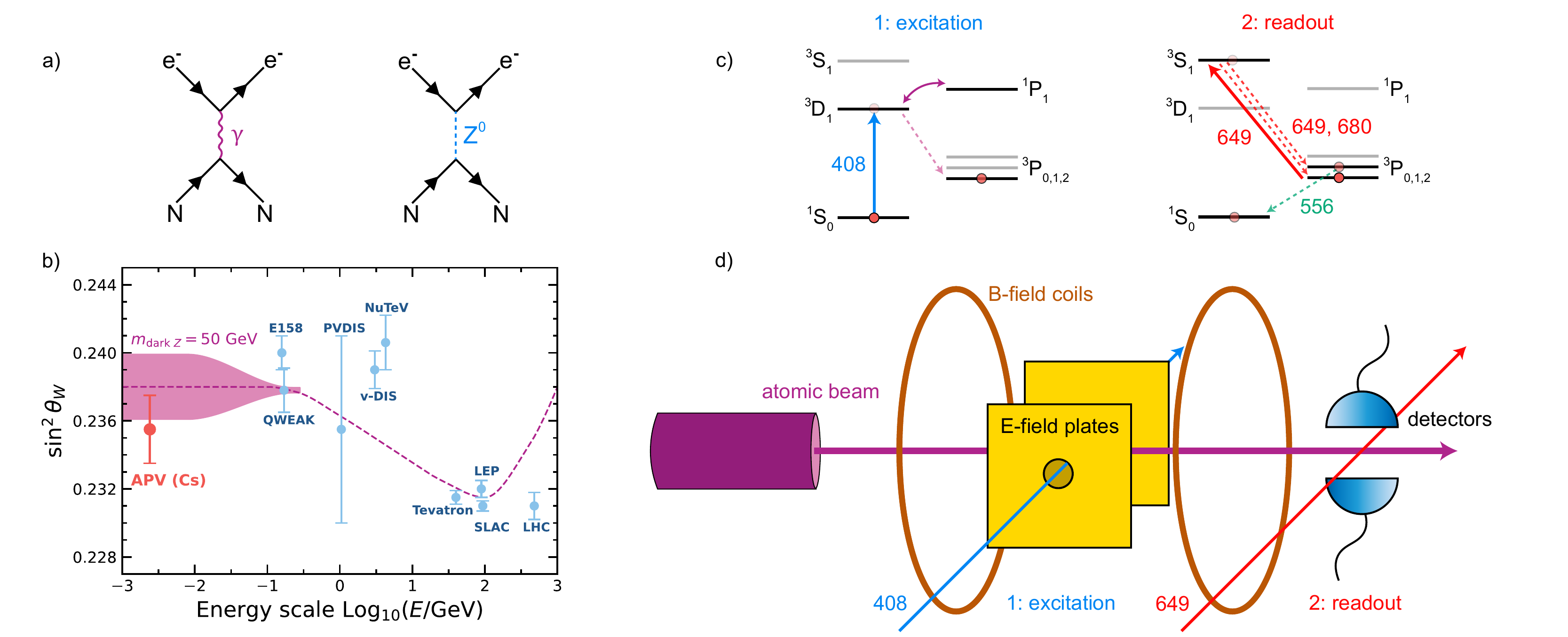}
\caption{
(a) Photon- and \(Z^0\)-mediated electron--nucleus interactions.  The weak neutral-current interaction violates parity and mixes opposite-parity atomic states.
(b) Standard Model tests versus momentum transfer, adapted from Refs.~\cite{Nanos2024,jefferson2018precision}.  APV provides the low-energy limit and probes the nuclear weak charge.
(c) Neutral-Yb APV level scheme. The \(408\,{\rm nm}\) \(6s^2\,{}^1S_0\rightarrow5d6s\,{}^3D_1\) transition is driven in external fields, producing Stark--PV interference (with $^3D_1$ and $^1P_1$ mixing). Atoms excited to \({}^3D_1\) partially decay to \(6s6p\,{}^3P_0\). The \({}^3P_0\) population is read out by \(649\,{\rm nm}\) excitation to \(6s7s\,{}^3S_1\) and detection of cascade fluorescence, including \(649\,{\rm nm}\), \(680\,{\rm nm}\), and \(556\,{\rm nm}\) photons.
(d) Atomic-beam APV apparatus, adapted from Ref.~\cite{Antypas2019PRA} (here specifically for Yb, similar for Cs): a thermal Yb beam intersects a \(408\,{\rm nm}\) enhancement cavity in crossed electric and magnetic fields, with field/polarization reversals used to isolate the parity-violating interference signal.
}
\label{fig:apv_intro}
\end{figure*}

The weak amplitude is far too small to observe directly as a transition rate.  Instead, it is measured by interference with a larger parity-conserving amplitude,
\begin{equation}
R\propto |A_{\rm PC}+A_{\rm PNC}|^2
\simeq
|A_{\rm PC}|^2+
2{\rm Re}(A_{\rm PC}^*A_{\rm PNC}).
\label{eq:interference}
\end{equation}
Here \(R\) is the transition rate, \(A_{\rm PC}\) is the parity-conserving reference amplitude, \(A_{\rm PNC}\) is the parity-nonconserving weak amplitude, and \({\rm Re}\) denotes the real part.  The term linear in \(A_{\rm PNC}\) changes sign under appropriate reversals of electric field, magnetic field, propagation direction, or polarization.  This reversal-odd interference term is the experimental APV signal.

For example, in neutral Yb, the relevant transition is the \(408\,{\rm nm}\) $^1S_0\rightarrow^3D_1$ line (\Cref{fig:apv_intro}c).  A DC electric field induces a Stark amplitude
\[
A_{\rm Stark}\propto\beta E,
\]
with the level $^1P_1$ (\Cref{fig:apv_intro}c), where \(E\) is the applied electric-field magnitude and \(\beta\) is the vector transition polarizability.  The weak interaction supplies the small parity-violating amplitude
\[
A_{\rm PV}\propto\zeta,
\]
where \(\zeta\) is the reduced weak-induced transition amplitude used in the Yb APV literature.  The measured ratio for \(^{174}{\rm Yb}\) is approximately
\[
\zeta/\beta\simeq -24\,{\rm mV/cm}.
\]
For a typical field \(E\sim1\,{\rm kV/cm}\), this corresponds to a weak-to-Stark amplitude ratio
\[
\frac{A_{\rm PV}}{A_{\rm Stark}}
\sim
\frac{\zeta}{\beta E}
\sim
2\times10^{-5},
\]
and therefore a reversal-odd rate contribution near the \(10^{-4}\) level.  This is why APV experiments require field reversals, harmonic detection, and detailed null tests \cite{Tsigutkin2009,Antypas2019PRA}.

The same interference idea can also be used coherently.  If an optical field couples through a parity-conserving Rabi frequency \(\Omega_{\rm PC}\) and a weak-induced Rabi frequency \(\Omega_{\rm PNC}\), so that
\[
\Omega=\Omega_{\rm PC}+\Omega_{\rm PNC},
\]
then an off-resonant dressing field detuned by \(\Delta\) produces the light shift
\[
\delta E
\simeq
\hbar\frac{|\Omega_{\rm PC}+\Omega_{\rm PNC}|^2}{4\Delta}.
\]
Here \(\hbar\) is the reduced Planck constant, \(\Delta\) is the laser detuning from the excited state, and \(\delta E\) is the induced energy shift.  Expanding this expression gives a parity-violating interference shift,
\[
\delta E_{\rm PNC}
\simeq
\hbar
\frac{2{\rm Re}(\Omega_{\rm PC}^*\Omega_{\rm PNC})}{4\Delta}.
\]
This shift can be read out as a Ramsey phase,
\[
\phi_{\rm PNC}=\delta E_{\rm PNC}\tau/\hbar,
\]
where \(\tau\) is the coherent interrogation time.  This is the basis of Fortson-type trapped-ion proposals and optical-lattice APV proposals \cite{Fortson1993,Kastberg2019,Craik2025}.

As a concrete example, the neutral-Yb experiment measures APV on the
\(408\,{\rm nm}\)
\[
6s^2\,{}^1S_0\rightarrow 5d6s\,{}^3D_1
\]
transition.  A thermal atomic beam crosses a power-enhanced standing wave of
\(408\,{\rm nm}\) light in a region with controlled static electric and magnetic
fields (\Cref{fig:apv_intro}d).  The electric field induces a Stark transition amplitude (mixing $^3D_1$ and $^1P_1$), while the weak
interaction induces a much smaller parity-violating amplitude.  Their
interference changes sign under appropriate reversals of the electric field,
magnetic field, or light polarization, allowing the APV contribution to be
separated from the much larger parity-conserving transition rate
\cite{Tsigutkin2009,Antypas2019NP,Antypas2019PRA}.

After excitation to \({}^3D_1\), a fraction of the atoms decay into the
metastable \(6s6p\,{}^3P_0\) state.  The experiment detects this population
rather than the short-lived \({}^3D_1\) state directly.  The \({}^3P_0\)
population is promoted to \(6s7s\,{}^3S_1\) with \(649\,{\rm nm}\) light, and
the resulting decay cascade produces fluorescence on several lines, including
\(649\,{\rm nm}\), \(680\,{\rm nm}\), and \(556\,{\rm nm}\).  By monitoring this
fluorescence while modulating and reversing the applied fields, the experiment
extracts a small reversal-odd APV signal from the transition probability.  This
is a rate/asymmetry measurement, not a coherent Ramsey measurement: the
short-lived \({}^3D_1\) state is populated and then decays, and the signal is
obtained from the accumulated fluorescence statistics.

\subsection{Experimental progress}

The benchmark single-isotope APV result is the Cs \(6S\rightarrow7S\) measurement, which reached sub-percent experimental precision and, with atomic theory, produced a low-energy determination of the weak charge and evidence for a nuclear anapole moment \cite{Wood1997,Dzuba2012Cs}.  Its strength is the combination of precision experiment and unusually mature many-body theory.

Yb provides a different advantage.  Tsigutkin \emph{et al.} observed a large APV effect in Yb on the \(408\,{\rm nm}\) transition, with a weak-induced amplitude much larger than in Cs \cite{Tsigutkin2009}.  Antypas \emph{et al.} then measured the isotopic variation of APV in the even isotopes
\[
^{170}{\rm Yb},\quad ^{172}{\rm Yb},\quad ^{174}{\rm Yb},\quad ^{176}{\rm Yb},
\]
confirming Standard Model weak-charge scaling and constraining new electron--nucleon interactions \cite{Antypas2019NP,Antypas2019PRA}.  The companion systematic study describes the harmonic-ratio method, field reversals, and error budget \cite{Antypas2019PRA}.

The power of isotope-chain APV is cancellation of the leading atomic factor:
\[
E_{\rm PNC}^{(A)}
=
K_{\rm el}Q_W(A)[1+\epsilon_A],
\]
so that
\begin{equation}
\frac{E_{\rm PNC}^{(A)}}{E_{\rm PNC}^{(A_{\rm ref})}}
\simeq
\frac{Q_W(A)}{Q_W(A_{\rm ref})}
[1+\epsilon_A-\epsilon_{A_{\rm ref}}].
\label{eq:isotope_ratio}
\end{equation}
Absolute many-body theory is replaced by smaller isotope-dependent corrections such as finite nuclear size, neutron skins, and residual electronic isotope shifts.

\subsection{Neutron skins and new forces}

Since \(Q_W\approx -N\), APV isotope ratios are sensitive to neutron distributions.  The difference between neutron and proton radii, the neutron skin, affects the overlap of the electron wavefunction with the weak charge distribution.  This makes isotope-chain APV complementary to nuclear scattering and isotope-shift measurements.

Isotope chains also probe new electron--neutron forces.  A new light mediator may generate an isotope-dependent contribution that does not follow Standard Model weak-charge scaling.  The neutral-Yb isotope-chain measurement already constrains such interactions \cite{Antypas2019NP,kniazev2024thesis}.  Modern isotope-shift and King-plot experiments provide related constraints, but APV has the distinctive feature of directly probing parity-violating electron--nucleus couplings.

\subsection{Trapped ions, lattices, and molecules}

Several proposals aim to convert APV from a beam rate measurement into coherent spectroscopy.  Fortson proposed measuring APV in a single trapped Ba\(^+\) ion using the \(S_{1/2}\rightarrow D_{3/2}\) quadrupole transition: the weak-induced electric-dipole (E1) amplitude interferes with the electric-quadrupole (E2) amplitude and produces a ground-state spin rotation \cite{Fortson1993}.  Optical-lattice proposals adapt this idea to large neutral ensembles, for example in Cs \cite{Kastberg2019}.  A recent entangled-ion proposal uses decoherence-free states to reject parity-conserving light shifts while retaining the PV vector light shift \cite{Craik2025}.

Yb\(^+\) is also promising (\Cref{fig:apv_array}).  The relevant transition is
\[
[4f^{14}]6s\,{}^2S_{1/2}
\rightarrow
[4f^{14}]5d\,{}^2D_{3/2}
\]
near \(435.5\,{\rm nm}\).  It is a narrow E2 clock transition; the weak interaction supplies a forbidden \(E1_{\rm PNC}\) amplitude.  
For \(A\simeq174\), the Yb\(^+\) PNC amplitude $\zeta$ is about \(6\times10^{-11}ea_0\) \cite{DzubaFlambaum2011},
roughly \(10\)--\(15\) times smaller than the measured neutral-Yb value
\(\zeta=8.7(1.4)\times10^{-10}ea_0\) on the \(408\,{\rm nm}\)
\({}^1S_0\rightarrow{}^3D_1\) line \cite{Antypas2019PRA}.  The tradeoff is coherence: neutral Yb has
the larger raw APV amplitude, while Yb\(^+\) has the clock-like
\(S_{1/2}\rightarrow D_{3/2}\) transition needed for Ramsey/cat metrology.

Molecules offer a complementary route, especially for NSD-PV.  Closely spaced opposite-parity levels can strongly enhance parity-violating mixing.  Molecular PV proposals and demonstrations using near-degenerate levels have been developed in Refs.~\cite{DeMille2008,Altuntas2018}, and new progress in laser-cooled molecule arrays \cite{Anderegg2019} opens potential many-body extensions. Recently, Gaul, Cong, and Budker showed that molecular parity-violation data can also be reinterpreted as constraints on new boson-mediated electron--nucleus interactions.  They used the measured parity violation in the hyperfine structure of \(^{138}{\rm Ba}^{19}{\rm F}\) from Ref.~\cite{Altuntas2018} together with electronic-structure calculations to bound previously unexplored axial-vector--vector nucleus--electron couplings.  The resulting constraints are comparable to those obtained from Cs APV, and the authors estimate that future cold heavy molecules such as \(^{137}{\rm BaF}\) or \(^{225}{\rm RaF}\) could improve the sensitivity to such interactions by at least two orders of magnitude for large mediator masses~\cite{gaul2026constraints}.

\section{Original proposal: APV as a quantum-metrology problem}

We model the isotope-chain APV signal as
\[
\omega_A=\Omega(q_A+\theta h_A),
\]
where \(q_A\) is the Standard Model weak-charge pattern, \(\Omega\) is an unknown common APV scale, \(\theta\) is a deviation from weak-charge scaling, and \(h_A\) is the assumed isotope-dependent pattern of a deviation (sign change). This deviation could represent a neutron-skin residual, a light-mediator contribution, or an isotope-dependent systematic. The key metrological question is: if \(N\) probes are available across the isotope chain, what quantum state best measures \(\theta\)?

Because \(\Omega\) is unknown, only the component of \(h_A\) orthogonal to the Standard Model pattern \(q_A\) is useful.  In a coherent Ramsey model, this defines a slope generator.  The information-optimal pure state is the equal superposition of the maximum- and minimum-eigenvalue branches of that generator.  We call this state the matched cross-isotope cat or just cross-isotope cat.  Its branches assign each isotope block a sign chosen to match the useful isotope-slope pattern.  Thus the state is not a product of independent isotope cats: it is one global cat matched to the single parameter \(\theta\).  For example, if the useful sign pattern for four isotopes is \(--++\), the cross-isotope cat has the schematic form
\[
\ket{\psi_{\rm cross}}
\sim
\frac{
\ket{--++}
+
e^{i\phi}\ket{++--}
}{\sqrt{2}},
\]
where each symbol represents an entire isotope subarray (\Cref{fig:apv_array}), see details in the Appendix.

A realistic cat should also reject common technical noise.  We therefore consider a decoherence-free subspace (DFS) cat, in which each isotope is encoded in two reversal channels with opposite APV sign but common magnetic and scalar light-shift noise \cite{Craik2025}. The two cat branches have the same phase under common noise and opposite phase under the APV generator.  This construction keeps the metrological structure of the cross-isotope cat while implementing the central idea of APV experiments: true APV changes sign under reversals, while ordinary shifts should not.  This is analogous to Fortson's trapped-ion idea, where the optical geometry is chosen so that the ordinary E2 light shift is common to the two ground-state Zeeman sublevels, while the PNC light shift changes sign and appears as a Larmor-frequency shift \cite{Fortson1993}.

\begin{figure}[htbp]
  \centering
  \includegraphics[width=8.5cm]{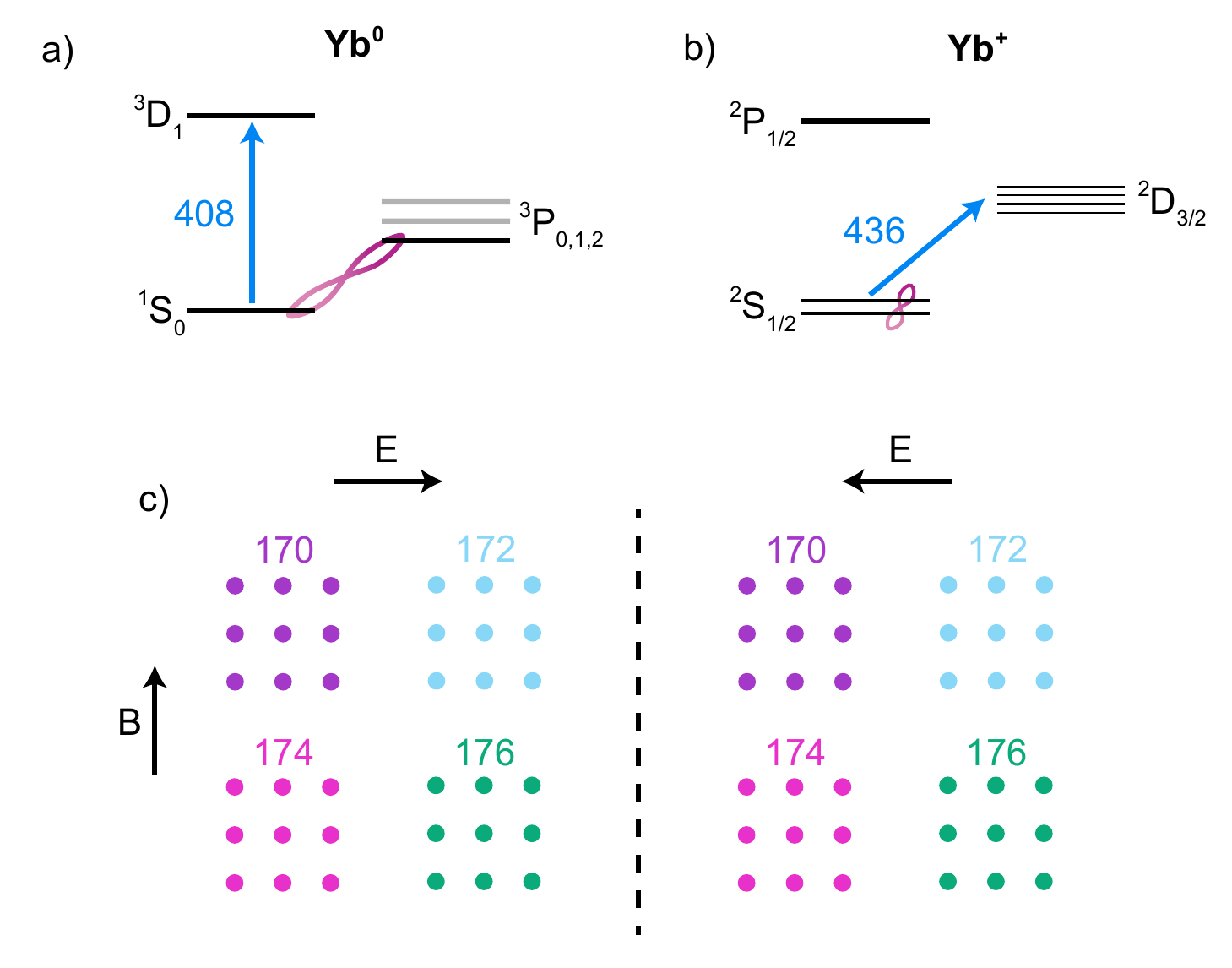}
\caption{
APV platforms for coherent metrology.
(a) Neutral Yb uses the demonstrated \(408\,{\rm nm}\)
\({}^1S_0\rightarrow{}^3D_1\) APV transition.  For even isotopes there is no
ground-state Zeeman qubit, so a Ramsey protocol would require mapping the
Stark--PV light shift onto a separate long-lived memory, such as the
\({}^1S_0-{}^3P_0\) optical clock transition.
(b) In Yb\(^+\), the \(435.5\text{--}436\,{\rm nm}\)
\({}^2S_{1/2}\rightarrow{}^2D_{3/2}\) quadrupole transition is naturally
compatible with Ramsey APV: the \({}^2S_{1/2}\) Zeeman doublet is the qubit, and
the weak-induced \(E1_{\rm PNC}\) amplitude interferes with the allowed \(E2\)
amplitude \cite{DzubaFlambaum2011,Porsev2012,kniazev2024thesis}.
(c) Array implementation. Isotope subarrays can be entangled into a
cross-isotope cat state. A second APV-reversed region, e.g.
\(\mathbf E\rightarrow-\mathbf E\), enables a DFS-type cat protocol in which common magnetic/trap shifts cancel while the APV phase adds \cite{Craik2025}.
}
\label{fig:apv_array}
\end{figure}

\section{Protocols compared in simulation}

\begin{figure*}[htbp]
  \centering
  \includegraphics[width=\textwidth]{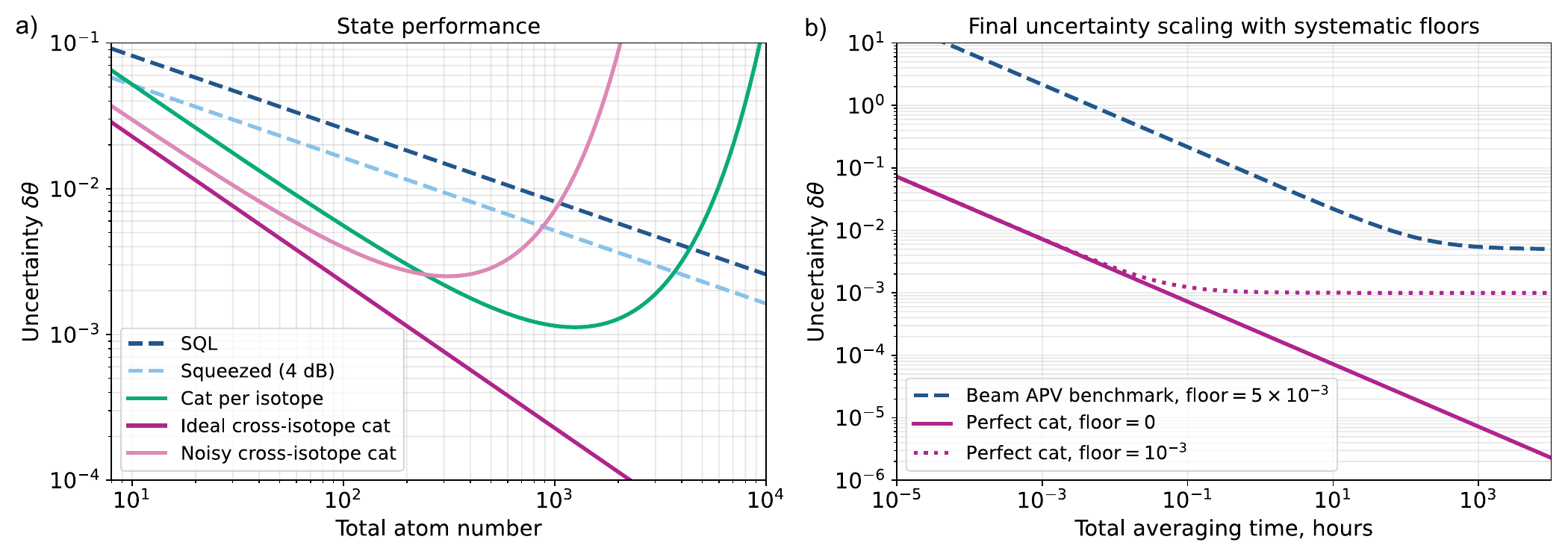}
\caption{
Projected metrological sensitivity to an APV isotope-chain deviation.
(a) Parameter uncertainty \(\delta\theta\) versus total atom number for \(1\,{\rm h}\) of averaging.
SQL measurements scale as \(N^{-1/2}\), while the ideal cross-isotope cat protocol reaches the Heisenberg limit \(\delta\theta\propto N^{-1}\).  Squeezed and same-isotope cat protocols give intermediate performance set by their assumed contrast and preparation errors.
(b) Uncertainty \(\delta\theta\) versus total averaging time for \(N=1000\).  Statistical
gains reduce the short-time uncertainty, but the total uncertainty saturates at
the systematic floor.
}
\label{fig:simulations}
\end{figure*}
We compare five strategies.  The first is the standard quantum limit (SQL): each isotope is measured independently and the resulting APV frequencies are combined in a classical fit.  The second is a spin-squeezed array, where each isotope subarray has reduced projection noise but remains more robust than a cat.  The third is a same-isotope cat protocol: each isotope is measured using its own Greenberger--Horne--Zeilinger (GHZ)-type cat state, and the isotope-chain slope is extracted classically.  The fourth is the ideal cross-isotope cat benchmark.  The fifth is a noisy cross-isotope cat model including finite contrast, gate infidelity, survival probability, and decoherence (\Cref{fig:simulations}).

The simulations show three main lessons.  First, in the statistics-limited regime, the ideal cross-isotope cat achieves the expected Heisenberg scaling and can reach a target slope sensitivity much faster than SQL.  Squeezing gives a smaller but more robust improvement.  Second, a noisy global cross-isotope cat can coincide with, beat or lose to the same-isotope cat strategy depending on the total atom number and gate fidelities.  The cross-isotope cat has a larger ideal slope gain, but it pays the contrast of one global cat over all atoms, while the same-isotope protocol uses several smaller cats, and is thus more robust to gate errors and decoherence in large arrays.  Third, the ultimate comparison with beam APV is controlled by systematics.  The final uncertainty should be written
\[
\delta\theta_{\rm tot}
=
\sqrt{\delta\theta_{\rm stat}^2+\sigma_{\theta,\rm sys}^2}.
\]
Entanglement reduces the statistical term, but it cannot average away a reversal-correlated APV systematic floor. The time scan (\Cref{fig:simulations}b) is plotted over a beam-experiment-scale averaging range: the
neutral-Yb isotope-chain measurement used 34 days of PV data acquisition,
corresponding to 420 hr of integration with about 62\% duty cycle
\cite{Antypas2019PRA}.  The \(5\times10^{-3}\) systematic floor is illustrative
and represents a present-day few-\(10^{-3}\) fractional APV precision scale,
not a directly measured value of the toy parameter \(\theta\).

This distinction is central.  A trapped array or ion chain can approach a target faster than an atomic beam if it has a smaller statistical coefficient.  However, if trap-induced or reversal-correlated systematics saturate at a higher floor, the array loses asymptotically.  The role of entanglement is therefore not to replace APV systematics; it is to make the statistical part less costly once the systematic channel is controlled.

\section{Feasibility}

Neutral Yb is experimentally compelling because the isotope-chain APV signal has already been measured.  The difficulty is coherent mapping.  The demonstrated \(408\,{\rm nm}\) APV transition uses the short-lived \({}^3D_1\) state, with lifetime about \(380\,{\rm ns}\) \cite{YbLifetime1996}.  This is fine for a beam rate measurement but problematic for a Ramsey light-shift protocol: spontaneous scattering destroys coherence.  One possible route is to use the even-Yb \({}^1S_0\rightarrow{}^3P_0\) clock transition as a memory and apply the \(408\,{\rm nm}\) Stark--PV interaction off resonantly to imprint a phase.  This is physically plausible, but not demonstrated.  The useful PV light shift is only \(\sim5\times10^{-5}\) of the ordinary Stark shift for typical Yb APV fields, so scattering and false reversal-odd light shifts are severe constraints.

Ions are more naturally suited to coherent APV.  In Ba\(^+\) and Yb\(^+\), the APV-sensitive \(S_{1/2}\rightarrow D_{3/2}\) transition is narrow and clock-like.  The APV signal appears as a spin-dependent light shift or Ramsey phase.  Even Yb\(^+\) isotope chains would probe NSI weak-charge scaling and could reduce absolute many-body theory dependence through isotope ratios.  The cost is smaller particle number and magnetic sensitivity of the Zeeman qubit.  This makes DFS reversal-pair protocols especially important.

The key feasibility issue in all platforms is the systematic floor.  Common-mode noise can be suppressed by reversal pairs, echo, and DFS encoding.  False APV signals cannot.  These include reversal-correlated electric-field errors, polarization leakage, vector or tensor light shifts, isotope-dependent trap shifts, and any ordinary shift that transforms like the weak signal.  The central experimental requirement is therefore to make the unwanted reversal-odd projection smaller than the weak-charge-slope precision one seeks.

\section{Conclusion and outlook}

We have reformulated isotope-chain APV as a quantum parameter-estimation problem.  In a coherent model, the optimal state for measuring a deviation from weak-charge scaling is a cross-isotope cat aligned with the isotope-slope generator. A DFS reversal-pair version preserves the same signal generator while cancelling common magnetic and light-shift noise.  Simulations show that entanglement can strongly reduce averaging time, but the final precision is set by APV-specific systematics.

The most plausible near-term path is probably not a single global cat.  More realistic milestones are squeezed isotope subarrays, small same-isotope cat states, simultaneous reversal channels, and ion-chain DFS protocols.  Neutral Yb offers a demonstrated large APV signal and clean even-isotope weak-charge chain, but coherent mapping is difficult.  Yb\(^+\) and Ba\(^+\) offer cleaner Ramsey physics, but smaller \(N\) and nontrivial theory/systematics.

Looking further ahead, molecular arrays may provide a new route.  Heavy molecules have closely spaced opposite-parity levels and large internal fields, making them attractive for PV.  Doyle and collaborators have demonstrated optical tweezer arrays of ultracold molecules \cite{Anderegg2019}, entanglement between CaF molecules \cite{Bao2023}, and polyatomic molecule arrays \cite{Vilas2024}.  One can imagine DFS molecular pairs in opposite electric-field orientations, cat states built from parity doublets, or hybrid atom--molecule arrays combining clean isotope ratios with enhanced NSD-PV sensitivity.  The long-term goal is not merely to entangle APV sensors, but to engineer a many-body state whose signal generator is the weak interaction and whose dominant noise generators are symmetry-forbidden.

\clearpage
\appendix

\section{Common-scale projection and the cross-isotope cat}

The isotope-chain model used in the simulations is
\[
\omega_A=\Omega(q_A+\theta h_A).
\]
Here \(A\) labels isotope, \(q_A=Q_W(A)/Q_W(A_{\rm ref})\) is the Standard Model weak-charge pattern, \(\Omega\) is an unknown common APV amplitude, and \(h_A\) is the assumed isotope-dependent pattern of a deviation.  If part of \(h_A\) is proportional to \(q_A\), it cannot be distinguished from changing the common scale \(\Omega\).  Therefore the useful direction is the component of \(h_A\) orthogonal to \(q_A\), weighted by atom number:
\[
h_{\perp,A}=h_A-\beta q_A,
\qquad
\beta=
\frac{\sum_A N_Ah_Aq_A}{\sum_A N_Aq_A^2}.
\]
This construction is independent of the arbitrary choice of reference isotope up to a relabeling of coordinates.  An overall sign flip of \(h_{\perp,A}\) simply swaps the two branches of the cat state.

For Ramsey time \(\tau\), the generator for \(\theta\) is
\[
G_\theta=
\pi\tau\Omega
\sum_A h_{\perp,A}
\sum_{j=1}^{N_A}\sigma_z^{(A,j)}.
\]
For a pure state,
\[
F_Q=4{\rm Var}(G_\theta).
\]
For any Hermitian generator, the maximum variance is obtained by the equal superposition of its maximum- and minimum-eigenvalue eigenstates.  Since each probe contributes \(\pm1\) to \(\sigma_z\), the maximum branch chooses the qubit sign of isotope \(A\) according to
\[
s_A={\rm sign}(h_{\perp,A}).
\]
Thus the information-optimal cross-isotope cat is
\[
\ket{\psi_{\rm cross}}
=
\frac{1}{\sqrt2}
\left[
\bigotimes_A\ket{s_A}^{\otimes N_A}
+
e^{i\phi}
\bigotimes_A\ket{-s_A}^{\otimes N_A}
\right].
\]
The two branches are product states over isotope groups, but their coherent superposition is a global inter-isotope cat.

For a four-isotope Yb toy model with the useful sign pattern \(--++\), this becomes schematically
\begin{multline*}
 \ket{\psi_{\rm cross}}
=
\frac{1}{\sqrt2}
\Big[
\ket{-}_{170}^{N_{170}}
\ket{-}_{172}^{N_{172}}
\ket{+}_{174}^{N_{174}}
\ket{+}_{176}^{N_{176}}
+ \\
e^{i\phi}
\ket{+}_{170}^{N_{170}}
\ket{+}_{172}^{N_{172}}
\ket{-}_{174}^{N_{174}}
\ket{-}_{176}^{N_{176}}
\Big].   
\end{multline*}

In the shorthand used in the main text this is
\[
\ket{\psi_{\rm cross}}
\sim
\frac{\ket{--++}+e^{i\phi}\ket{++--}}{\sqrt2}.
\]
The eigenvalue separation is
\[
\Delta\lambda
=
2\pi\tau\Omega
\sum_A N_A|h_{\perp,A}|,
\]
so after \(RT_{\rm avg}\) repetitions,
\[
\delta\theta_{\rm cross}
=
\frac{1}{
2\pi\tau\Omega
\left(\sum_A N_A|h_{\perp,A}|\right)
\sqrt{RT_{\rm avg}}
}.
\]
This is the Heisenberg-scaling benchmark.

\section{DFS cat and Zeeman-noise cancellation}

For each isotope, introduce two reversal channels \((A,+)\) and \((A,-)\).  These could be two spatial subarrays, two ion-chain regions, or two optical standing-wave phases.  They are engineered so that the APV phase has opposite sign,
\[
\omega_{\rm APV}^{(A,+)}=+\omega_A^{\rm APV},
\qquad
\omega_{\rm APV}^{(A,-)}=-\omega_A^{\rm APV},
\]
while ordinary magnetic, laser, and scalar light-shift noise is common:
\[
\omega_{\rm noise}^{(A,+)}\simeq \omega_{\rm noise}^{(A,-)}.
\]
The DFS cat is
\begin{multline*}
\ket{\psi_{\rm DFS}}
=
\frac{1}{\sqrt2}
\Big[
\bigotimes_A
\ket{s_A}_{A,+}^{\otimes N_A}
\ket{-s_A}_{A,-}^{\otimes N_A}
\\
+
e^{i\phi}
\bigotimes_A
\ket{-s_A}_{A,+}^{\otimes N_A}
\ket{s_A}_{A,-}^{\otimes N_A}
\Big].
\end{multline*}
Common noise couples to
\[
S_z^{A,+}+S_z^{A,-},
\]
whereas APV couples to
\[
S_z^{A,+}-S_z^{A,-}.
\]
Both cat branches have the same eigenvalue under the common-noise operator but opposite eigenvalues under the APV operator.  Thus common Zeeman and scalar light-shift phases cancel, while the APV phase adds.

This is closely related to the logic of Fortson's proposal.  In Fortson's Ba\(^+\) scheme, the optical geometry is chosen so that the ordinary E2 light shift is independent of the ground-state spin \(m=\pm1/2\), while the PNC interference shift has opposite sign for the two spin states.  The measured signal is a shift of the Larmor precession frequency.  In the many-body DFS version, the same idea is implemented by entangling reversal channels: common Zeeman phase is rejected, while the PV phase survives.

The DFS construction does not cancel systematics that share the APV reversal signature.  Reversal-correlated electric-field errors, polarization leakage, tensor light shifts, or isotope-dependent trap shifts remain as the floor \(\sigma_{\theta,\rm sys}\).

\section{States and sensitivity models used in simulation}

The simulations compare the following protocols.

\paragraph{SQL.}
Each isotope is measured with independent probes.  The single-isotope uncertainty is
\[
\delta\omega_A^{\rm SQL}
=
\frac{1}{2\pi C\tau\sqrt{N_A RT_{\rm avg}}},
\]
and the isotope-chain parameter \(\theta\) is obtained from a classical Fisher fit.

\paragraph{Squeezed array.}
Each isotope subarray is spin squeezed.  The uncertainty is
\[
\delta\omega_A^{\rm sq}
=
\xi\,\delta\omega_A^{\rm SQL},
\]
where \(\xi=10^{-G_{\rm dB}/20}\).  In the plotted examples we use a net squeezing gain of \(4\,{\rm dB}\), interpreted as an achieved metrological gain.

\paragraph{Same-isotope cat.}
Each isotope is measured with its own GHZ-type cat,
\[
\ket{\psi_A}
=
\frac{
\ket{+}_A^{\otimes N_A}
+
e^{i\phi}\ket{-}_A^{\otimes N_A}
}{\sqrt2}.
\]
The isotope measurements are then combined classically.  The single-isotope sensitivity is
\[
\delta\omega_A^{\rm cat}
=
\frac{1}{2\pi C_{N_A}\tau N_A\sqrt{RT_{\rm avg}}}.
\]

\paragraph{Cross-isotope cat.}
The optimal benchmark is the global cat over isotope subarrays,
\[
\ket{\psi_{\rm cross}}
\sim
\frac{\ket{--++}+e^{i\phi}\ket{++--}}{\sqrt2}
\]
for the illustrative four-isotope sign pattern.  This directly measures the isotope-chain slope generator.

\paragraph{Noisy cross-isotope cat.}
For ordinary cats we use the contrast model
\[
C_N=
C_0F_1^{n_1}F_2^{n_2}
p_{\rm surv}^N
e^{-N\tau/T_2}.
\]
For the metrological plot in Fig.~\ref{fig:simulations}(a), typical optimistic parameters are
\[
\tau=1\,{\rm s},\qquad T_2\gg\tau,
\]
\[
1-F_1=10^{-4},\qquad 1-F_2=10^{-3},
\]
with \(T_{\rm avg}=1\,{\rm h}\).  These values are not a claim about a realized APV array; they are a clean hardware-level benchmark showing the scaling of different quantum states.

\paragraph{DFS cross-isotope cat.}
A DFS cat should be modeled with separate local and residual differential noise,
\[
C_N^{\rm DFS}
\sim
C_0F_1^{n_1}F_2^{n_2}p_{\rm surv}^N
e^{-N\tau/T_{2,\rm local}}
e^{-\tau/T_{2,\rm diff}}.
\]
Here \(T_{2,\rm diff}\) is the residual coherence time after common-mode cancellation.  This form expresses the fact that common Zeeman noise does not scale as catastrophically with \(N\) if the state is encoded in reversal pairs. We do not show these simulations on the plot since the plot assumes no magnetic field common noise, but the model can be used for more experiment-tailored calculations.

\section{Time scan and systematic floor}

For the time-scan plot in Fig.~\ref{fig:simulations}(b), we fix the number of probes to
\[
N=1000
\]
and compare how the uncertainty improves with total averaging time.  If a protocol has statistical uncertainty \(\delta\theta_{\rm stat}(T_0)\) after time \(T_0\), then
\[
\delta\theta_{\rm stat}(T)
=
\delta\theta_{\rm stat}(T_0)\sqrt{\frac{T_0}{T}}.
\]
Including a non-averaging APV floor gives
\[
\delta\theta(T)
=
\sqrt{
\delta\theta_{\rm stat}^2(T_0)\frac{T_0}{T}
+
\sigma_{\theta,\rm sys}^2
}.
\]
Thus entanglement reduces the coefficient of \(1/\sqrt{T}\), but cannot beat the asymptotic floor.  This is why the array curve can initially fall faster than the beam benchmark but still lose asymptotically if its APV systematic floor is higher.

\section{Neutral-Yb and Yb\(^+\) mapping details}

For neutral even Yb, the demonstrated APV transition is
\[
{}^1S_0\rightarrow{}^3D_1
\]
at \(408\,{\rm nm}\).  The even isotopes have \(I=0\), and the ground state has \(J=F=0\), so there is no ground-state Zeeman qubit.  A coherent Ramsey protocol would therefore require a separate memory, such as the
\[
{}^1S_0\rightarrow{}^3P_0
\]
clock transition.  The \(408\,{\rm nm}\) field could then be used off resonantly to imprint a Stark--PV light shift on the \({}^1S_0\) branch of the clock superposition.  This is physically allowed, but not yet demonstrated.  The difficulty is that the \({}^3D_1\) state is short-lived and the useful PV light shift is a tiny reversal-odd part of a much larger ordinary Stark shift.

For Yb\(^+\), the coherent scheme is more natural.  The relevant transition is
\[
{}^2S_{1/2}\rightarrow{}^2D_{3/2}
\]
near \(435.5\,{\rm nm}\).  The \({}^2S_{1/2}\) ground state has a Zeeman doublet and can be used as a Ramsey qubit.  The \({}^2D_{3/2}\) level has four Zeeman sublevels, which must be included in a real calculation.  One can either resolve selected Zeeman components using a bias magnetic field and polarization selection, or follow the Fortson-style approach where the optical geometry sums over the allowed \(D_{3/2}\) paths in such a way that the ordinary E2 light shift is common while the PNC light shift is spin-odd.

The spin-rotation picture and Ramsey spectroscopy describe the same Hamiltonian measurement.  If
\[
H_{\rm PNC}=\frac{\hbar\omega_{\rm PNC}}{2}\sigma_z,
\]
then
\[
\frac{\ket{\uparrow}+\ket{\downarrow}}{\sqrt2}
\rightarrow
\frac{
e^{-i\omega_{\rm PNC}\tau/2}\ket{\uparrow}
+
e^{i\omega_{\rm PNC}\tau/2}\ket{\downarrow}
}{\sqrt2}.
\]
The spin has rotated by \(\omega_{\rm PNC}\tau\), which is also the Ramsey phase.  A same-isotope cat amplifies this phase by \(N_A\), while a DFS state cancels common magnetic noise.

\bibliography{sample}

\end{document}